\documentclass{aa}
\usepackage[varg]{txfonts}
\usepackage[utf8]{inputenc}
\usepackage{amsmath}
\usepackage{graphicx}
\usepackage{float}
\usepackage{enumerate}
\usepackage{bm}
\usepackage{siunitx}
\usepackage{mhchem}
\usepackage{xcolor}
\usepackage[allcolors=blue,colorlinks=true]{hyperref}
\pdfoutput=1

\bibpunct{(}{)}{;}{a}{}{,} 
\newcommand\pd[2]{\dfrac{\partial #1}{\partial #2}}
\newcommand\pdd[2]{\dfrac{\partial^2 #1}{\partial #2^2}}

\DeclareMathOperator{\lapl}{\nabla^2}

\title{The effect of internal gravity waves on cloud evolution in sub-stellar atmospheres}

\institute{Division of Computing and Mathematics, Abertay University, Kydd Building, Dundee DD1 1HG, UK\\\email{1303985@abertay.ac.uk, c.stark@abertay.ac.uk}\label{abertay}
\and
Mathematics, School of Science \& Engineering, University of Dundee, Nethergate, Dundee DD1 4HN, U.K.\label{dundee}
\and
Atmospheric, Oceanic and Planetary Physics, Department of Physics, University of Oxford, Oxford OX1 3PU, UK\label{oxford}
}

\author{A.~Parent\inst{\ref{abertay}}
\and R.~E.~Falconer\inst{\ref{abertay}}
\and E.~K.~H.~Lee\inst{\ref{oxford}}
\and K.~A.~Meyer\inst{\ref{dundee}}
\and C.~R.~Stark\inst{\ref{abertay}}}

\titlerunning{Sub-stellar internal gravity waves}
\authorrunning{A. Parent et al.}
\date{Received ... / Accepted ...}

\abstract
{Sub-stellar objects exhibit photometric variability, which is believed to be caused by a number of processes, such as magnetically-driven spots or inhomogeneous cloud coverage. Recent sub-stellar models have shown that turbulent flows and waves, including internal gravity waves, may play an important role in cloud evolution.}
{The aim of this paper is to investigate the effect of internal gravity waves on dust nucleation and dust growth, and whether observations of the resulting cloud structures could be used to recover atmospheric density information.}
{For a simplified atmosphere in two dimensions, we numerically solved the governing fluid equations to simulate the effect on dust nucleation and mantle growth as a result of the passage of an internal gravity wave. Furthermore, we derived an expression that relates the properties of the wave-induced cloud structures to observable parameters in order to deduce the atmospheric density.}
{Numerical simulations show that the density, pressure, and temperature variations caused by gravity waves lead to an increase of the dust nucleation rate by up to a factor $20$, and an increase of the dust mantle growth rate by up to a factor $1.6$, compared to their equilibrium values. Through an exploration of the wider sub-stellar parameter space, we show that in absolute terms, the increase in dust nucleation due to internal gravity waves is stronger in cooler (T dwarfs) and \ce{TiO2}-rich sub-stellar atmospheres. The relative increase, however, is greater in warm (L dwarf) and \ce{TiO2}-poor atmospheres due to conditions that are less suited for efficient nucleation at equilibrium. These variations lead to banded areas in which dust formation is much more pronounced, similar to the cloud structures observed on Earth.}
{We show that internal gravity waves propagating in the atmosphere of sub-stellar objects can produce banded clouds structures similar to that observed on Earth. We propose a method with which potential observations of banded clouds could be used to estimate the atmospheric density of sub-stellar objects.}


\begin{document}
\maketitle

\section{Introduction}\label{sec:intro}

Brown dwarfs are low-mass, sub-stellar objects below the hydrogen burning limit, with masses between $13\,M_\textnormal{Jup}$ and $70\,M_\textnormal{Jup}$. As a consequence, their atmospheres are sufficiently cool for the formation of dust clouds. The clouds are observed through their effect on spectral features and their association with infrared spectroscopic variability, which is believed to be caused by patchy clouds. Numerous observations \citep[see for example][]{Buenzli_2014} show that a large portion of known brown dwarfs exhibit photometric variability. According to \citet{Biller_2017}, over 10\% of known brown dwarfs show a variation of 1\% or more, and over 50\% exhibit a variation of 0.1\% to 0.5\% or more.

Explaining spectral variability is necessary to understanding the L/T transition \citep{Vos_2019}. The L and T components of the Luhman~16AB system, for example, show vastly different patterns \citep{Gillon_2013}: Luhman~16B, a T dwarf, exhibits strong, fast-changing periodic spectral variations, while the L dwarf Luhman~16A exhibits no periodic pattern. A model from \citet{Marley_2008} and \citet{Marley_2010} proposes the sinking of parts of the cloud deck, creating thinner, patchy clouds as an explanation for the change in variability patterns around the L/T transition. A study by \citet{Stark_2015} proposes the electrostatic disruption of charged cloud particles as a mechanism through which inhomogeneous coverage could be caused. While inhomogeneous dust cloud coverage \citep{Helling_2014} is believed to be the main cause for brown dwarf variability, other theories, such as temperature variations \citep{Robinson_2014} or fingering convection \citep{Tremblin_2016}, propose cloud-free models as an explanation.

Internal gravity waves have been simulated in main sequence stars \citep{Alvan_2014}. Internal gravity waves triggered by fingering convection have been modelled in objects ranging from main-sequence stars to brown dwarfs \citep{Garaud_2015}, and they can reach wavelengths much larger than the source perturbation. Simulations of atmosphere patches by \citet{Freytag_2010} show that internal gravity waves are also present in sub-stellar atmospheres, triggered by downdrafts caused by convection patterns, and they are theorised to be one of the main phenomena responsible for transporting dust in the upper atmospheric layers.

Further characterising the impact of gravity waves on dust cloud evolution can advance the understanding of cloud structures in brown dwarfs. \citet{Helling_2001} showed that higher frequency acoustic waves, triggered by turbulent flow in brown dwarfs atmospheres, can have a strong impact on cloud formation: by carrying lower temperature perturbations, the passage of waves can temporarily create favourable conditions for dust nucleation in otherwise dust-hostile environments, leading to the formation of dust over time.

Internal gravity waves are a type of fluid wave that occurs in atmospheres and oceans. Their defining characteristic is that gravity, in the form of buoyancy, is the restoring force that allows disturbances to propagate. Internal gravity waves can be observed in the Earth's atmosphere through their effect on clouds. On Earth, a wave cloud is formed from the passage of an internal gravity wave, triggered by stable air flowing over relief. The vertical displacement of the air forces it to oscillate as the buoyancy force tries to restore equilibrium. As the wave propagates, at the wave peaks the displaced air rises and cools resulting in water vapour condensing, forming droplets and clouds; at the wave troughs, the clouds evaporate due to adiabatic heating, leading to clouds that have a distinct banded structure.

In the case of a gas giant planet or brown dwarf, dust clouds are formed instead of water clouds but an analogous process can occur. In this context, wave clouds can be induced as a result of external fluid motion triggering turbulent flow (such as fingering convection in deeper layers of the atmosphere), whereas gas flow over relief would be expected to be the main cause in the case of a rocky terrestrial exoplanet \citep{Roeten_2019}. In a sub-stellar atmosphere, oscillating parcels of gas can trigger the nucleation of seed particles and enhanced surface mantle growth, forming banded cloud structures. The nucleation rate is a function of density of the nucleating species and the atmospheric temperature. If the passage of the internal gravity wave perturbs the local thermodynamic structure of the atmosphere it can give enhanced nucleation in localised regions.

The aim of this paper is to investigate and characterise the effect of internal gravity waves on the evolution of dust clouds in the atmospheres of sub-stellar objects and its consequences for cloud variability. This paper presents a novel mechanism for potentially diagnosing the gas density of sub-stellar atmospheres from observations of the resulting cloud structures formed from the passage of internal gravity waves. In Sect.~\ref{sec:model} the basic atmospheric model of internal gravity waves, nucleation and mantle growth is described; in Sect~\ref{sec:numerical} the numerical methods used to simulate internal gravity waves are presented; in Sect.~\ref{sec:numerical.results} the results of the simulations are presented and discussed; Sect.~\ref{sec:discussion} summarises and discusses the consequences of the results including a possible way of connecting observations to the wave dispersion relation to diagnose the atmospheric density.

\section{Sub-stellar internal gravity waves}
\label{sec:model}
For a vertical slice of a sub-stellar atmosphere in hydrostatic equilibrium, the coupled equations of fluid dynamics governing the evolution of the fluid velocity $\vec{u}$, the fluid density $\rho$, and the pressure $p$, of an atmospheric parcel, under the influence of gravity $\vec{g}$ are:
\begin{align}
&\pd{\rho}{t}+\nabla\cdot(\rho\vec{u})=0, \label{eqn:continuity} \\
&\rho\left[\pd{\vec{u}}{t}+(\vec{u}\cdot\nabla)\vec{u}\right]=-\nabla p-\rho \vec{g}, \label{eqn:momentum} \\
&p\rho^{-\gamma_{a}}=\textnormal{const.}, \label{eqn:state}
\end{align}
where $\gamma_{a}$ is the ratio of specific heats (for a diatomic gas $\gamma=7/5$). In static equilibrium $\vec{u}_0=0$ and $\partial/\partial t=0$, giving the following equilibrium relationship:
\begin{align}
&\nabla p_0 = -\rho_{0} \vec{g},
\end{align}
where the subscript `$0$' denotes an equilibrium quantity. For simplicity, in order to capture the fundamental physics, this paper focuses on the effect of gravity waves in the linear regime. We can linearise Eqs.~(\ref{eqn:continuity})-(\ref{eqn:state}) by decomposing each variable $Q$ into its equilibrium and perturbed value, so that $Q=Q_0 + Q_1$. In the non-linear regime $Q_0 \gg Q_1$, and powers of $Q_1$ higher than $1$ can be discarded. For clarity, the subscript `$1$' is omitted in further equations. Linearisation yields the final system of fluid equations:
\begin{align}
\pd{\rho}{t}&=-\nabla\cdot(\rho_{0}\vec{u}), \\
\pd{\vec{u}}{t}&=-\frac{\nabla p}{\rho_{0}}-\frac{\rho\vec{g}}{\rho_{0}}.
\end{align}

To model internal gravity waves, where we deal with incompressible flows $\nabla\cdot\vec{u}=0$, we adopted a vorticity-stream function formulation by introducing the vorticity $\vec{\zeta}$ and stream function $\vec{\psi}$ defined by,
\begin{align}
\vec{\zeta} &= \nabla\times\vec{u}, \label{eqn:vort-def}\\
\vec{u} &= \nabla\times\vec{\psi}. \label{eqn:strf-def}
\end{align}

Therefore, the governing fluid equations for incompressible flows in the linear regime become,
\begin{align}
\pd{\rho}{t}&=-(\nabla\times\vec{\psi})\cdot\nabla\rho_{0}, \\
\pd{\vec{\zeta}}{t}&=-\nabla\times\left(\frac{\rho\vec{g}}{\rho_{0}}\right), \\
\vec{\zeta}&=\nabla\times(\nabla\times\vec{\psi}).
\end{align}
where the baroclinic term ($\nabla\rho\times\nabla p/\rho^{2}$) vanishes since the propagation of internal gravity waves is considered to be an adiabatic process. In an atmospheric vertical plane $(x,y)$:
\begin{align}
\vec{u} &= \left(u_x,u_y,0\right) = \left(\pd{\psi}{y},-\pd{\psi}{x},0\right), \\
\vec{\psi} &= \left(0,0,\psi\right), \\
\vec{\zeta} &= \left(0,0,\pdd{\psi}{x}+\pdd{\psi}{y}\right), \\
\vec{g} &= (0,g,0).
\end{align}

This yields the system of equations
\begin{align}
\pd{\zeta}{t} &= -\frac{g}{\rho_0} \pd{\rho}{x}, \label{eqn:zeta-l} \\
\pd{\rho}{t} &= \pd{\rho_0}{y} \pd{\psi}{x}, \label{eqn:rho-l} \\
\zeta &= -\nabla^{2}\psi, \label{eqn:strf-l}
\end{align}
where
\begin{equation}
\lapl=\left(\pdd{}{x}+\pdd{}{y}\right).
\end{equation}

To obtain the gravity waves' dispersion relation, Eq.~(\ref{eqn:strf-l}) is derived with respect to time, and Eq.~(\ref{eqn:zeta-l}) substituted for $\partial\zeta/\partial t$:
\begin{equation}
\pd{\zeta}{t} = -\lapl\left[\pd{\psi}{t}\right] = -\dfrac{g}{\rho_0} \pd{\rho}{x}. \label{eqn:disp-1}
\end{equation}

Differentiating Eq.~(\ref{eqn:disp-1}) with respect to time, and then substituting Eq.~(\ref{eqn:rho-l}) for $\partial \rho/\partial t$ yields:
\begin{align}
\lapl\left[\pdd{\psi}{t}\right] = \dfrac{g}{\rho_0} \pd{}{x}\left(\pd{\psi}{x}\pd{\rho_0}{y}\right).
\end{align}

At equilibrium, $\rho_0$ does not vary along $x$, Therefore:
\begin{align}
\lapl\left[\pdd{\psi}{t}\right] = \dfrac{g}{\rho_0}\pd{\rho_0}{y}\pdd{\psi}{x}= -N^2\pdd{\psi}{x}, \label{eqn:disp-2}
\end{align}
where
\begin{align}
N = \sqrt{-\frac{g}{\rho_0}\pd{\rho_0}{y}}, \label{eqn:bv-freq}
\end{align}
is the Brunt-V\"ais\"al\"a buoyancy frequency. In the case of a uniformly stratified atmosphere, assuming a solution of the form $\psi \approx \exp{[-i(\omega t + k_{x}x+k_{y}y)]}$ the dispersion relation for internal gravity waves becomes~\citep{Sutherland_2010,Vallis_2017},
\begin{align}
\omega &= \sqrt{\frac{N^2k_{x}^2}{k_{x}^2+k_{y}^2}}, \label{eqn:disp}\\
       &= N\cos{\theta}. \label{eqn:disp-theta}
\end{align}

Therefore, when the atmospheric density (or equivalently velocity) is perturbed, corresponding oscillations in $\rho$, $\vec{\psi}$, and $\vec{\zeta}$ are triggered, that occur at the Brunt-V\"ais\"al\"a buoyancy frequency $N$. As the wave propagates through the atmosphere, the density variations can affect the resulting nucleation and mantle growth rates.

\subsection{Dust nucleation}
\label{sec:model-dust}
To quantify the impact of passing waves on dust formation, we used the equation of modified classical nucleation theory presented by \citet{Gail_1984, Helling_2001}, which defines the nucleation rate $J_*$, the number of nucleating centres formed per second per unit volume [\SI{}{\per\m\cubed\per\s}], as:
\begin{align}
&J_{*} = \frac{n_x}{\tau}Z\exp\left[(N_* - 1)\ln{S}-\left(\dfrac{T_\theta}{T}\right)\dfrac{N_* - 1}{(N_* - 1)^{1/3}}\right], \label{eqn:nucleation}
\end{align}
where $T$ is the temperature; $\tau$ is the seed growth time scale for the gaseous nucleation species $x$; $N_{*}$ is the size of the critical cluster; $n_{x}$ is the number density of the nucleating species; $Z$ is the Zeldovich factor; $S$ is the supersaturation ratio; defined as follows:
\begin{align}
&\tau= n_x v_{\textnormal{rel},x} N_*^{2/3}A_{0}, \\
&N_* = 1 + \left(\dfrac{2T_{\theta}}{3T\ln{S}}\right)^3, \\
&Z =\left[\dfrac{T_{\theta}}{9\pi T} \left( N_* - 1\right)^{4/3}\right]^{1/2},\\
&S= \frac{p_{x}}{p_{\textnormal{sat},x}},
\end{align}
where,
\begin{align}
&\ln{(p_{\textnormal{sat},x})}= 35.8027-\frac{74734.7}{T}~[\text{dyn}],~~~T\in[500,2500]~\textnormal{K}, \\
&v_{\textnormal{rel},x} \approx \sqrt{ \dfrac{k_B T}{2 \pi m_x}}, \\
&T_{\theta}=\frac{4\pi r_{0}^{2}\sigma}{k_{B}}, \\
&r_{0}=\left(\frac{3Am_{p}}{4\pi\rho_{m}}\right)^{1/3},
\end{align}
and $p_{x}$ is the partial pressure of the nucleating species $x$;  $p_{\textnormal{sat},x}$ is the saturation vapour pressure of the nucleating species $x$; $\sigma$ is the surface tension of the nucleating species; $r_{0}$ is the hypothetical monomer radius; $A_{0}=4 \pi r_0^2$ is the hypothetical monomer surface area; $m_{x}$ is the mass of a monomer particle; $m_{p}$ is the proton mass; $A$ is the atomic weight of the monomer; $\rho_{m}$ is the dust material density; and  $v_\textnormal{rel}$ is the thermodynamic equilibrium velocity for the nucleating species studied ($\ce{TiO2}$ for this paper).

For this paper, we assumed a constant value for the surface tension of \ce{TiO2}, $\sigma_{\ce{TiO2}}=\SI{0.618}{\J\per\m\squared}$ (see \citet{Helling_2001, Lee_2015}), and a density $\rho_{m}\approx \SI{4230}{\kg\per\m\cubed}$ for \ce{TiO2}. In the context of an internal gravity wave propagating through a sub-stellar atmosphere, the wave perturbs the local atmospheric gas density and hence the temperature (via $T\rho^{1-\gamma_{a}}=\textnormal{const.}$) in an adiabatic process. As a result, the passage of the wave perturbs the nucleation rate $J_{*}$.

\subsection{Mantle growth}

Once nucleation has established a material surface onto which material can accumulate, dust growth occurs via gas-phase surface chemistry (Eq.~(24) in \citet{Helling_2006}). We consider a spherical dust grain of radius $a$, of mass $m_{d}$, and let $n$ be the number density of the gas phase. The dust grain absorbs gas molecules at a rate $\alpha\pi a^{2}n \langle v\rangle$, where $\langle v\rangle$ is the mean gas molecular speed and $\alpha$ is the sticking probability that a molecule is absorbed (the sticking factor). Therefore, the mass of the dust grain $m_{d}$ evolves in time as
\begin{equation}
\frac{\textnormal{d}m_{d}}{\textnormal{d}t}=\alpha\pi a^{2}nm \langle v\rangle=\alpha\pi a^{2}\rho \langle v\rangle,
\end{equation}
where $\rho =nm$ is the gas mass density. The mass of a dust grain can be written as $m_{d}=\frac{4}{3}\pi a^{3}\rho_{m}$, where $\rho_{m}$ is assumed to be constant. Therefore, the time evolution of the radius of a dust grain is
\begin{equation}
\frac{\textnormal{d}a}{\textnormal{d}t}=\frac{\alpha \rho\langle v\rangle}{4\rho_{m}}=\gamma, \label{eqn:growth-rate}
\end{equation}
where $\gamma$ is the growth rate from absorption in units of $[\mbox{m}\mbox{s}^{-1}]$. Eq.~(\ref{eqn:growth-rate}) is the archetypal equation describing the absorption of material onto the surface of a dust grain. It is consistent with the dust growth equations presented in \citet{Helling_2001}, albeit in a much simplified form but still encapsulating the fundamental underlying physics. Furthermore, Eq.~(\ref{eqn:growth-rate}) is also consistent with mantle growth via ion accretion when dust grains are immersed in a plasma (Eq.~(18) in \citet{Stark_2018}). Without loss of generality, to investigate the effect of internal gravity waves on the mantle growth rate we simplify the expression by introducing $\rho_s$, the density of the gas-phase accreting species, defined as $\rho_s=f_s\,\rho$, where $0 \leq f_s \leq 1$ is the fraction of the surrounding gas composed of the accreting species,
\begin{equation}
     \gamma = \dfrac{f_s\,\rho v_{\textnormal{rel},s}}{4 \rho_{m}}, \label{eq:mantle-growth}
\end{equation}
where we have set $\alpha = 1$ to obtain the optimal growth rate; and $\langle v\rangle=v_{\textnormal{rel},s}$. Introducing $f_{s}$ allows us to generalise the effect of different gas-phase species, with varying relative abundances in the gas-phase, participating in surface chemistry leading to mantle growth.

\section{Numerical simulations}
\label{sec:numerical}
\subsection{Model equations}
The linearised governing equations can be cast in non-dimensional form, defining
\begin{align}
\tau&=t/T, \\
\xi&=x/L, \\
\lambda&=y/L, \\
a&=\zeta T, \\
b&=\rho L^{3}/M, \\
c&=\psi T/L^{2}, \\
\beta&=g T^{2}/L,
\end{align}
where $L$, $T$, and $M$ are characteristic values of length, time, and mass respectively. Therefore, in non-dimensional form, Eqs.~(\ref{eqn:zeta-l})-(\ref{eqn:strf-l}) become
\begin{align}
\pd{a}{\tau}&=-\dfrac{\beta}{a_{0}}\pd{b}{\xi} \label{eqn:zeta-ndl}, \\
\pd{b}{\tau}&=\pd{a_{o}}{\lambda}\pd{c}{\xi} \label{eqn:rho-ndl}, \\
a&=-\pdd{c}{\xi}-\pdd{c}{\lambda} \label{eqn:strf-ndl}.
\end{align}

Casting the model equations in non-dimensional form lets us observe the characteristic behaviour of internal gravity waves without loss of generality.

\subsection{Methods}
\label{sec:numerical.methods}

To solve the system of fluid equations (\ref{eqn:zeta-ndl}) - (\ref{eqn:strf-ndl}) numerically, we used a combination of the leapfrog method for Eq.~(\ref{eqn:zeta-ndl}) and Eq.~(\ref{eqn:rho-ndl}), and Successive Over-Relaxation (SOR) for Eq.~(\ref{eqn:strf-ndl}) \citep{Vetterling_1989, Mittal_2014}. SOR requires that the value of $\zeta$ from Eq.~(\ref{eqn:strf-l}) be known at the boundaries of the domain. In order to minimise the artefacts caused by boundaries, we used a domain large enough that over short timescales, the waves' perturbations do not reach the boundaries. Additionally, values of $\rho$, $\psi$, $\zeta$, and their spatial first-order derivatives were interpolated using third-order polynomial interpolation. The internal gravity waves were triggered by creating a Gaussian density perturbation initial condition in the centre of the numerical domain, that was modulated in time by a sine wave with a period equal to a multiple of the local buoyancy frequency,
\begin{equation}
    \rho_{1} =\rho_{A} \sin\left(\omega \tau\right)\exp\left(-\varsigma r^{2}\right), \label{eqn:gaussian}
\end{equation}
where $\rho_{A}$ is the maximum amplitude of the perturbation; $\varsigma$ is the spread parameter; and $r$ is the distance to the centre of the numerical domain. The resulting wave solutions were used to calculate the nucleation (Eq.~(\ref{eqn:nucleation})) and dust mantle growth (Eq.~(\ref{eq:mantle-growth})) rates as a result of the propagating internal gravity waves. The simulation parameters are presented in Table~\ref{table:sim.params}.
\begin{table}
\caption{Parameters used for numerical simulations}
\label{table:sim.params}
\centering
\begin{tabular}{l l l}
\hline\hline
Parameter & Value & \\
\hline
$n_x$ & $2000$ & no. of grid points along $x$ \\
$n_y$ & $1000$ & no. of grid points along $y$ \\
$h$ & $\num{1e-3}$ & spacial mesh increment \\
$\omega$ & $[0.25,~1.0]$ & driving perturbation frequency \\
$n_\tau$ & 50 & total no. of time steps \\
$d\tau$ & $T_\omega/n_\tau$ & temporal mesh increment \\
\hline
\end{tabular}
\end{table}

\subsection{Sub-stellar atmospheric model}
\label{sec:numerical.atmo}

\begin{figure}[h]
    \centering
    \resizebox{\hsize}{!}{\includegraphics{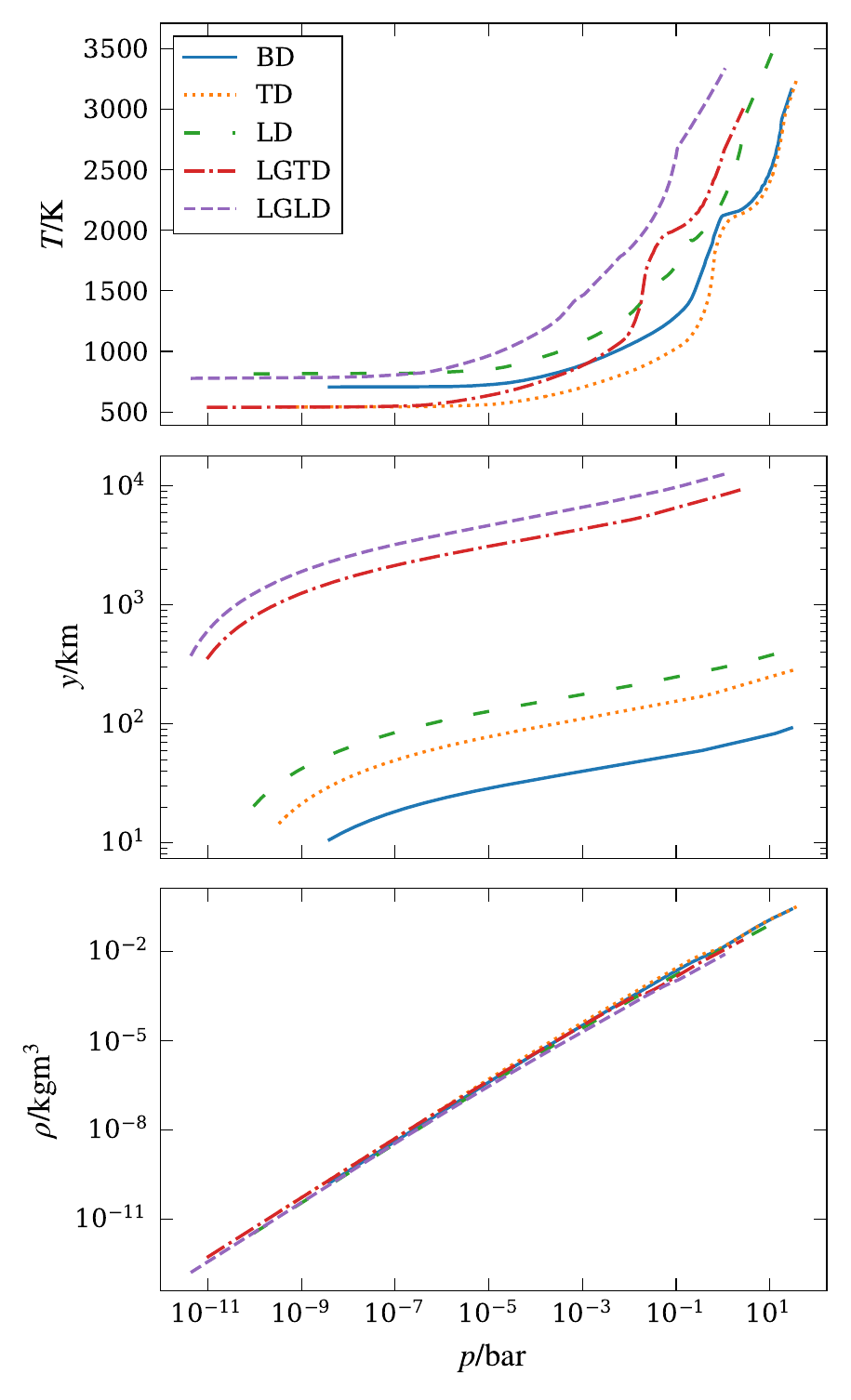}}
    \caption{Atmospheric diagram ($p, T$; top panel), ($p, y$; middle panel), and ($p, \rho$; bottom panel) for the \textsc{Drift-Phoenix} brown dwarf atmosphere model used in numerical simulations (BD: $T_\textnormal{eff}=\SI{1500}{\K}, \log{g}=5.0$, published by \citet{Stark_2013} and \citet{Rodriguez_2018}). Additional profiles for typical (LD, TD: $\log(g)=4.5$) and low-gravity (LGLD, LGTD: $\log(g)=3.0$) L dwarf ($T_\textnormal{eff}=\SI{2000}{\K}$) and T dwarf ($T_\textnormal{eff}=\SI{1200}{\K}$) models are shown for context (see also Table~\ref{table:atmo.models}.)}
    \label{fig:atmo.profile}
\end{figure}

The aim of this paper was to investigate the effects of internal gravity waves on the evolution of dust clouds in sub-stellar atmospheres. To this end, we considered a brown dwarf atmosphere characterised by $T_\textnormal{eff}=\SI{1500}{\K}$ and $\log{g}=5.0$ (see Fig. \ref{fig:atmo.profile}) as an exemplar sub-stellar atmosphere, and we used data published in \citet{Stark_2013} as input for our numerical simulations. Figure~\ref{fig:atmo.profile} and Table~\ref{table:atmo.models} also give the profiles of typical and low-gravity L and T dwarfs for context, to place the simulations in a wider range of brown dwarfs examples. The atmospheric data was generated by the \textsc{Drift-Phoenix} model atmosphere and cloud formation code \citep{Hauschildt_1999, Helling_2004,Helling_2006,Witte_2009,Witte_2011}. We note that the atmospheric extension $y$ (Fig.~\ref{fig:atmo.profile}; middle panel) is measured from the top of the atmosphere.

\begin{figure}[h]
    \centering
    \resizebox{\hsize}{!}{\includegraphics{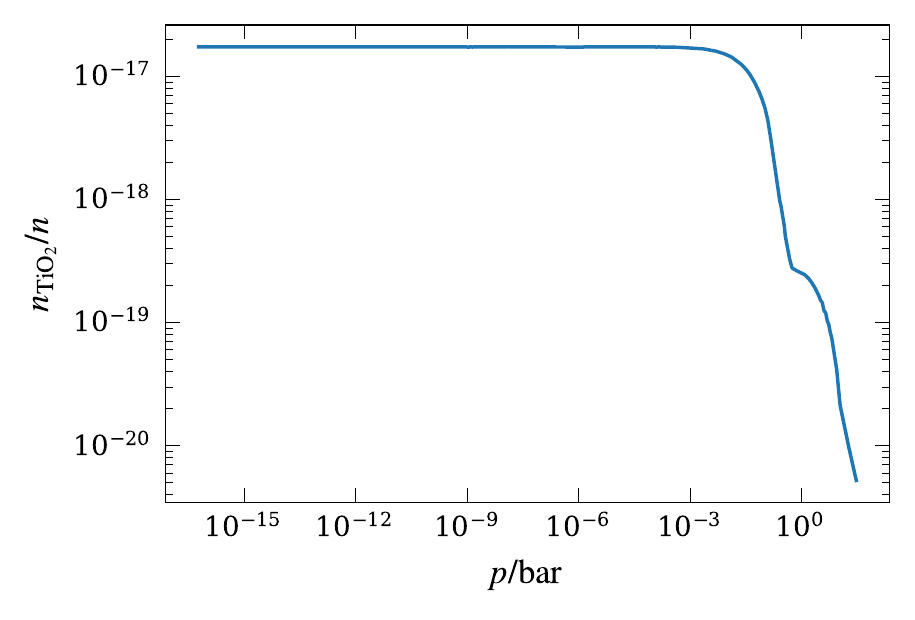}}
    \caption{Plot of the partial number density of $\ce{TiO2}$ at equilibrium for the model used in simulations ($\log{g}=5.0, T_\textnormal{eff}=\SI{1500}{\K}$). In the range of pressures used in the simulations (\SI{e-5}{\bar}), $n_{\ce{TiO2}}/n$ is constant around $10^{-17}$.}
    \label{fig:atmo.ntio2}
\end{figure}

The atmosphere model used as input is one-dimensional; for this study, we assumed a horizontally uniform atmosphere at equilibrium, and expanded the model to two dimensions. We chose the characteristic length $L$ (see Sect.~\ref{sec:numerical.methods}) to be the height of the simulation domain ($10^{4}$m). The numerical domain was defined by the geometric extension $y \in [\SI{25}{\km}, \SI{35}{\km}]$; the pressure $p_0 \in [\SI{3e-5}{\bar}, \SI{3e-3}{\bar}]$; the temperature $T_0 \in [\SI{700}{\K}, \SI{850}{\K}]$; and the density $\rho_0 \in [\SI{e-7}{\kg\per\m\cubed}, \SI{e-5}{\kg\per\m\cubed}]$. For these conditions, $n_{\ce{TiO2}}/n = \textnormal{const.} \approx 10^{-17}$ (see Fig.~\ref{fig:atmo.ntio2}). The perturbation length scale was of the order of \SI{e3}{\m}.


\begin{table}
\caption{Models presented in Figs.~\ref{fig:atmo.profile} and \ref{fig:atmo.period}}
\label{table:atmo.models}
\centering
\begin{tabular}{l l l l}
\hline\hline
Key & $T_\textnormal{eff}$ & $\log{g}$ & description \\
\hline
BD & \SI{1500}{\K} & $5.0$ & Brown Dwarf (simulations) \\
TD & \SI{1200}{\K} & $4.5$ & Typical T dwarf \\
LD & \SI{2000}{\K} & $4.5$ & Typical L dwarf \\
LGTD & \SI{1200}{\K} & $3.0$ & Low-gravity T dwarf \\
LGLD & \SI{2000}{\K} & $3.0$ & Low-gravity L dwarf \\
\hline
\end{tabular}
\end{table}

\section{Results}
\label{sec:numerical.results}

\begin{figure}[h]
    \centering
    \resizebox{\hsize}{!}{\includegraphics{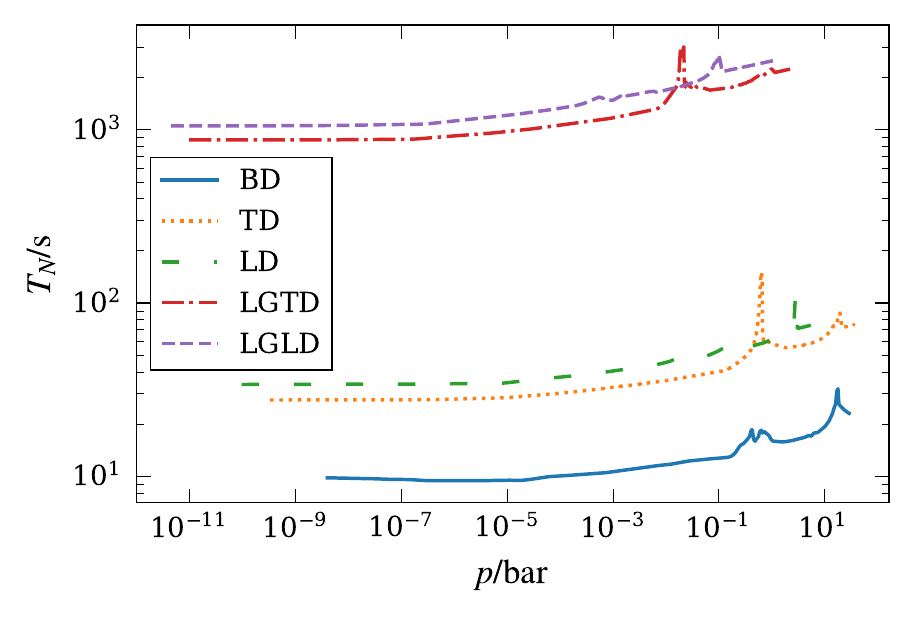}}
    \caption{Buoyancy period $T_{N}=2\pi/N$ for the \textsc{Drift-Phoenix} brown dwarf ($\log{g}=5.0, T_\textnormal{eff}=\SI{1500}{\K}$) atmosphere model used in numerical simulations, as well as typical and low-gravity L and T dwarfs (see Fig.~\ref{fig:atmo.profile}, Table~\ref{table:atmo.models}). The range of periods for higher surface gravity levels (\SIrange{e1}{e2}{\s}) is consistent with the atmospheric models and simulation results published by \citet{Freytag_2010}. In lower-density brown dwarfs, the buoyancy period rises to the order of \SI{e3}{\s}.}
    \label{fig:atmo.period}
\end{figure}

In a stratified sub-stellar atmosphere, perturbing the background density by vertically displacing a fluid parcel triggers vertical oscillations as the buoyancy force tries to restore equilibrium. The resulting density variations propagate through the atmosphere at the Brunt-V\"ais\"al\"a buoyancy frequency. Figure~\ref{fig:atmo.period} shows the buoyancy period $T_{N}=2\pi/N$ as a function of pressure $p$ for the sub-stellar atmosphere characterised by $T_\textnormal{eff}=\SI{1500}{\K}$ and $\log{g}=5.0$. The buoyancy period is of the order of \SI{10}{\s} across the extent of the atmosphere, with the maximum period occurring at high atmospheric pressures. Also shown in Fig.~\ref{fig:atmo.period}, for context, are the buoyancy periods profiles of typical L and T dwarfs (periods of the order of \SIrange{10}{100}{\s}), and low-gravity L and T dwarfs (longer buoyancy periods, of the order of \SI{1000}{\s}). In comparison, \citet{Buenzli_2014} observe spectroscopic variations on timescales of \SIrange{100}{1000}{\s}. The key parameter distinguishing between the buoyancy period for different atmospheric models is the surface gravity; since $T_{N}\propto g^{-1/2}$, objects with lower surface gravity will have a longer buoyancy periods. In the case of a neutrally stratified atmosphere (i.e. $N=0$), the potential temperature is constant with atmospheric height and internal gravity waves cannot propagate \citep{Sutherland_2010}. This scenario is expected deeper in the atmosphere at higher gas pressures \citep[for example, see][]{Tremblin_2015,Tremblin_2019}.

The wavelength of the internal gravity wave is determined by the spatial length-scale of the instigating perturbation driving the oscillation, and consequently sets the speed of propagation. If the initial density variation is driven by an external source, such as convective motions or large-scale turbulent motions, the frequency of the wave is set by the frequency of the source for frequencies below the buoyancy frequency. In sub-stellar atmospheres the length scale of convective motions deep in the atmosphere is related to the atmospheric pressure scale height by a factor between 1 and $10^{2}$ \citep{Marley_2015,Tremblin_2019} -- for example, $l_{\textnormal{conv}} \approx \SI{e3}{\km}$ in \citet{Freytag_2010} -- varying with timescales of the order of \SI{e3}{\s} \citep{Tremblin_2019}.

\begin{figure}[h]
    \centering
    \resizebox{\hsize}{!}{\includegraphics{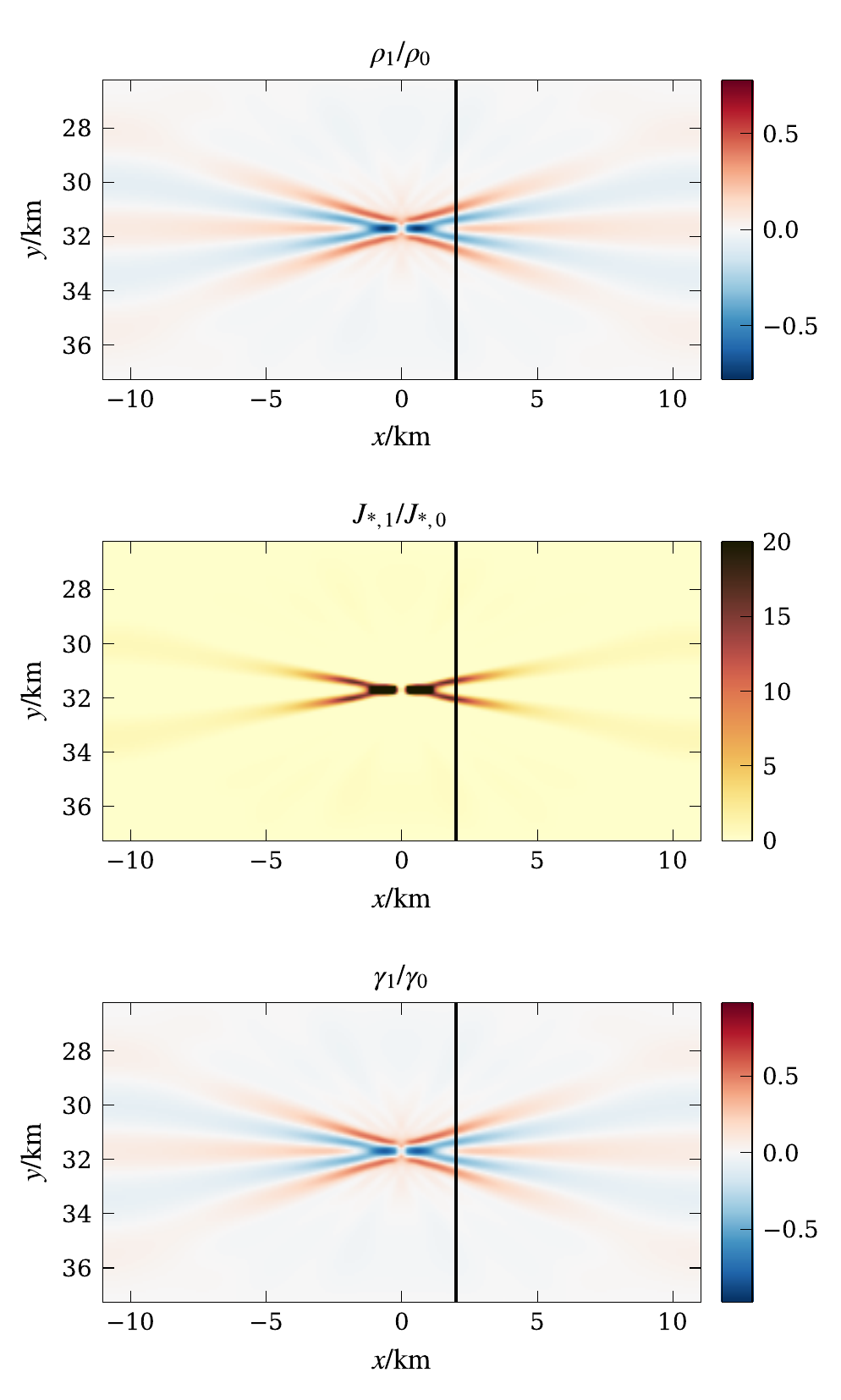}}
    \caption{2D map of $\rho_{1}/\rho_{0}$ (top panel), $J_{*,1}/J_{*,0}$ (middle panel) and $\gamma_{1}/\gamma_{0}$ (bottom panel) after one oscillation period. For this example, waves are driven at $\omega=0.25 N$, and the amplitude of the driving perturbation $\rho_{A} = 0.2~\rho_{0}$. The vertical line shows the location of the slice used to measure the wave response, at a distance $X(\omega)$ from the centre of the domain so that $X(\omega) = X_{\textnormal{ref}}\omega_{\textnormal{ref}}/\omega$.}
    \label{fig:demo}
\end{figure}

Figure~\ref{fig:demo} shows the characteristic St Andrews cross pattern emanating from density perturbation located at the centre of the numerical domain, for the example driving frequency $\omega=0.25 N$ and a driving perturbation amplitude $\rho_\textnormal{A} = 0.2 \rho_0$. In an adiabatic process the density variations are accompanied by sympathetic variations in temperature ($T\propto \rho^{\gamma_{a}-1}=\rho^{2/5}$) that go on to affect the local nucleation rate and surface mantle growth rate. Figure~\ref{fig:demo} shows that areas where $\rho$ is reduced by the passage of internal gravity waves are the areas with the largest increase of dust nucleation. The number density $n_x$ of the nucleating species is a component of the total gas density $\rho=\sum_{s}m_{s}n_{s}$ (See Fig.~\ref{fig:atmo.ntio2}, showing the relationship between $n$ and $n_{\ce{TiO2}}$). Variations of $\rho$ propagated by passing gravity waves are therefore reflected in the partial density of the nucleating species $n_{\ce{TiO2}}$, which in turns leads to an adiabatic change in temperature both of the total gas phase and its components. The primary effect that drives the variation in nucleation rate is the temperature variation: as the density increases (decreases) there is a corresponding adiabatic increase (decrease) in the temperature. Increasing (decreasing) the temperature of the nucleating species in this way, decreases (increases) the supersaturation ratio, $S$. Efficient nucleation is only possible for temperatures $T_{0}$ such that $S(T_{0})\gg 1$ \citep{Helling_2001}; therefore, increasing (decreasing) the temperature can yield a corresponding decrease (increase) in the supersaturation ratio, $S$, and hence a decrease (increase) in the nucleation rate. The nucleation rate is strongly dependent on the supersaturation ratio and hence the temperature; therefore, local over-densities in the atmosphere as a result of an internal gravity wave can decrease the local nucleation rate; whereas, local under-densities can increase the nucleation rate.

If there is an established particulate onto which material from the atmospheric gas can be absorbed, the growth rate can be also be affected by the passage of an internal gravity wave. In this scenario, an increase (decrease) in the local density, then the number of gas particles passing through a target area per unit time also increases (decreases), and so does the number of interactions per unit time that occurs between the seed particulates and the atmospheric species.

\begin{figure}[h]
    \centering
    \resizebox{\hsize}{!}{\includegraphics{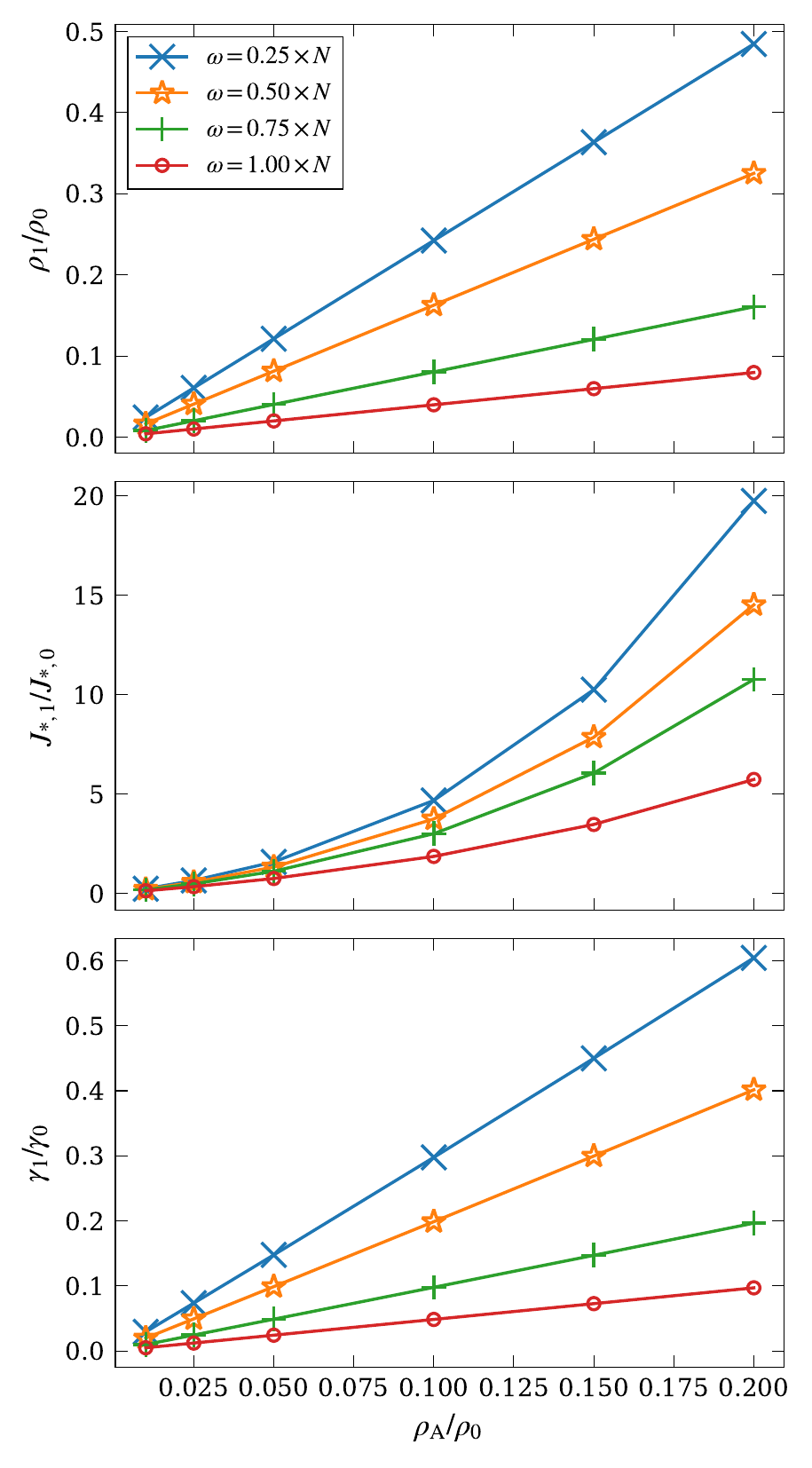}}
    \caption{Plots of $\rho_{1}/\rho_{0}$ (top panel), $J_{*,1}/J_{*,0}$ (middle panel), and $\gamma_{1}/\gamma_{0}$ (bottom panel) after one period of an internal gravity wave, as a function of the amplitude of the density perturbation used to drive the waves. The nucleation plot shows a strong non-linear increase in response to increased perturbation amplitude, more pronounced with low driving frequencies. The impact on dust growth is much weaker, and growth increases with the driving frequency. The measurements are taken as the maximum values along a vertical slice of the numerical domain, located at a distance $X(\omega)$ from the centre of the domain so that $X(\omega) = X_{\textnormal{ref}}\omega_{\textnormal{ref}}/\omega$.}
    \label{fig:amp-dust}
\end{figure}

To quantify the impact of the waves on dust nucleation and growth, we ran simulations for a range of driving frequencies and perturbation amplitudes. The density, nucleation and growth responses are presented in Fig. \ref{fig:amp-dust}. To obtain results comparable across varying frequencies, the values used for plotting were measured for $\tau=T_{\omega}$, where $T_{\omega}$ is the period of the driving oscillations, and normalised to their values at static equilibrium in the centre of the numerical domain. In order to obtain results comparable across a range of frequencies, the measurements were taken as the maximum values for $\rho_{1}$, $J_{*,1}$, and $\gamma_{1}$ along a vertical slice of the numerical domain, located at a distance $X(\omega)$ from the centre of the domain so that $X(\omega) = X_{\textnormal{ref}}\omega_{\textnormal{ref}}/\omega$, where $X_{\textnormal{ref}}$ is the location of the slice picked for a reference case at $\omega_{\textnormal{ref}}$.

As the perturbation amplitude increases the density amplitude of the resulting wave response linearly increases by up to a factor $1.5$, consistent with the linear regime assumed (top panel, Fig.~\ref{fig:amp-dust}). In contrast, the resulting nucleation rate (middle panel, Fig.~\ref{fig:amp-dust}) increases non-linearly with the perturbation amplitude by up to a factor $20$ in the most favourable scenario ($\omega=0.25 N$, $\rho_{A}=0.1\rho_{0}$). This is a result of the complex non-linear dependence of the nucleation rate (Eq.~(\ref{eqn:nucleation})), and not as a consequence of a non-linear evolution of the internal gravity wave, since our simulations are conducted in the linear regime only. This implies that if the internal gravity wave were to evolve non-linearly, the corresponding nucleation rate could exhibit an enhanced non-linear response, giving greater nucleation rate values. A similar assertion can be made regarding the mantle growth rate (bottom panel, Fig.~\ref{fig:amp-dust}): the mantle growth rate increases by up to a factor $1.6$ as a non-linear function of the perturbed density, $\gamma\propto \rho T^{1/2}=\rho^{(\gamma_{a}-1)/2}$; however, the non-linear dependence is less pronounced than that for the nucleation rate. The normalised growth rate $\gamma_{1}/\gamma_{0}$ is independent of the fraction of the surrounding gas composed of the accreting species $f_{s}$. Varying the spatial scale of the initial density perturbation does not affect the amplitude of the wave response.

When $\omega>N$, internal gravity waves cannot propagate since the system cannot respond quick enough to the imposed driven perturbation and the waves are evanescent. When $\omega\leq N$, internal gravity waves can freely propagate, where the interplay between the driving frequency and the local buoyancy frequency results in greater response amplitudes for lower driving frequencies than for frequencies approaching the natural buoyancy frequency of the system (see Fig.~\ref{fig:amp-dust}). As the generated wave propagates away from the oscillation source, it encounters regions of differing background density and hence local buoyancy frequency. In response, the wave amplitude, speed, and wavelength change in harmony to conserve wave energy. For example, if the wave propagates into regions of lower-density (higher-density), the amplitude of the wave increases (decreases), the wave speed increases (decreases), and the wavelength decreases (increases) in sympathy. If the wave encounters a region where its frequency is greater than the local buoyancy frequency, the wave ceases to propagate.

\begin{figure}[h]
    \centering
    \resizebox{\hsize}{!}{\includegraphics{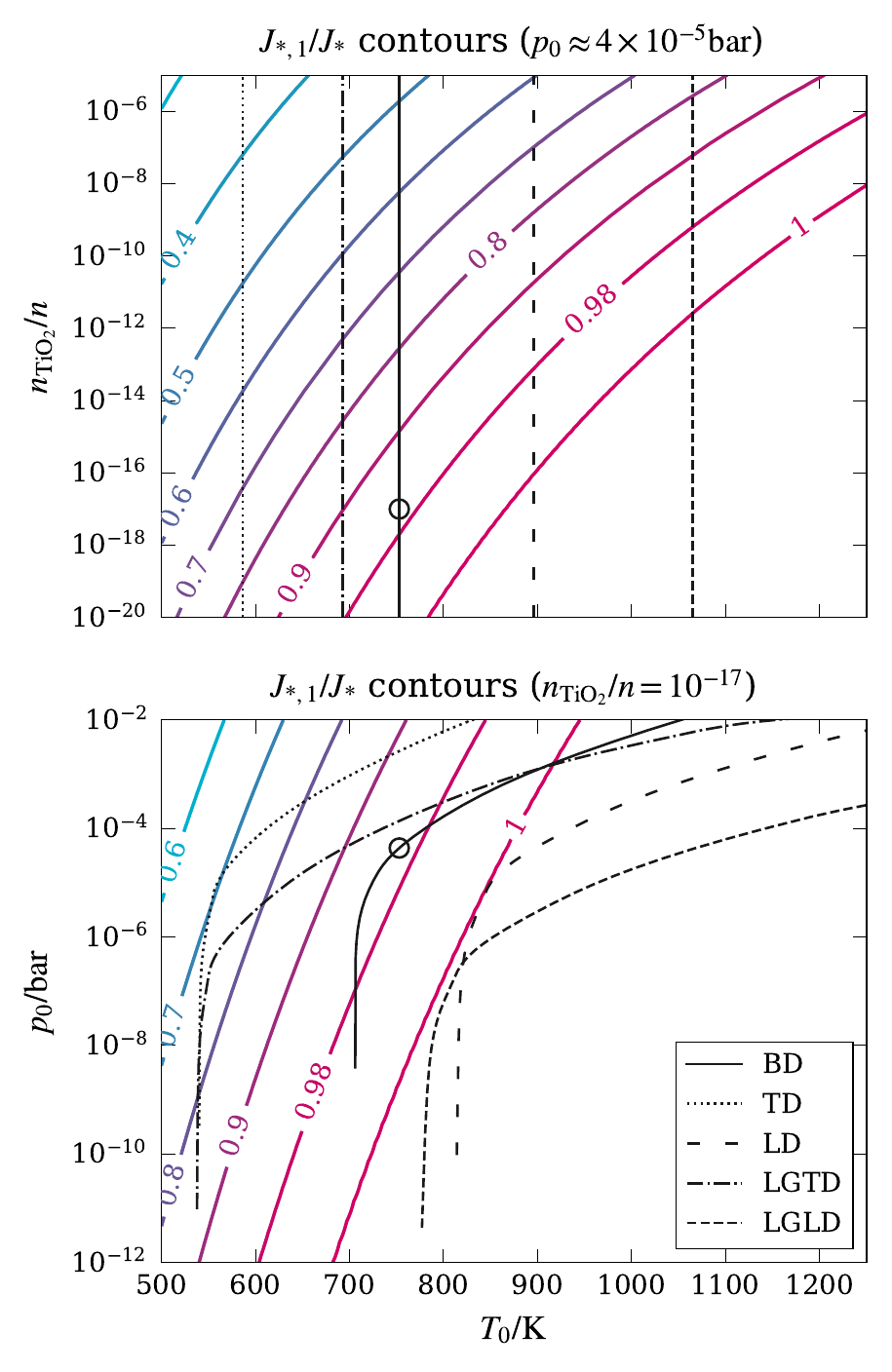}}
    \caption{Contour of the change in nucleation rate $J_{*,1}$ (normalised to the total nucleation rate $J_{*}$), as a function of equilibrium temperature $T_0$, $n_{\ce{TiO2}}/n$ (top panel) and equilibrium gas pressure $p_0$ (bottom panel). The atmospheric profiles for the brown dwarf model used for simulations (BD: $T_\textnormal{eff}=\SI{1500}{\K},\log g = 5.0$), as well as typical (LD,TD) and low-gravity (LGLD, LGTD) L and T dwarfs are plotted for context (see model details in Table~\ref{table:atmo.models}; Fig.~\ref{fig:atmo.profile}). The circle on both panels indicate the conditions present in the model used for simulations at the centre of the domain.}
    \label{fig:param-space}
\end{figure}

The numerical simulations presented in Fig.~\ref{fig:amp-dust} are normalised to equilibrium reference values, aiding in their generalisation to other brown dwarf and exoplanet models. For example, the density response (top plot, Fig~\ref{fig:amp-dust}) and the dust growth rate response (bottom plot, Fig.~\ref{fig:amp-dust}) are indicative of the typical response expected in atmospheric models beyond the exemplar presented. In contrast, the nucleation rate response is more complex and is driven by three main parameters: the ambient gas temperature $T$, the total gas pressure $p$, and the number density $n_x$ of the nucleating species $x$. These interdependent parameters depend upon a number factors, including the chemical composition of the atmosphere and the chemical processes involved \citep[eg. see][]{Helling_2017,Lee_2015, Lee_2018}. To investigate the nucleation rate response, we explored the parameter space of the key atmospheric variables, in order to contextualise the results beyond the exemplar atmospheric model considered. Varying $T_{\textnormal{eff}}$ and $\log{g}$ explicitly instead can obfuscate the physical picture leading to the underlying cause of the nucleation rate variations and can be misleading when dealing with microphysical processes. Figure~\ref{fig:param-space} shows the contour of the change in nucleation rate $J_{*,1}$ (normalised to the total nucleation rate $J_{*}$) for a range of background temperatures $T_0 \in [\SI{500}{\K}, \SI{1200}{\K}]$, pressures $p_0 \in [\SI{e-15}{\bar}, \SI{e-2}{\bar}]$, and $n_{\ce{TiO2}}/n \in [10^{-20}, 10^{-5}]$ to cover a wide range of values from contemporary models (see Fig.~4, \citet{Lee_2015}; Fig.~4, \citet{Helling_2008}; Fig.~2 \citet{Stark_2013}).

For the top plot of Fig.~\ref{fig:param-space}, we set $p_0 \approx 4 \times \SI{e-5}{\bar}$ (taken from the centre of the domain shown in Fig.~\ref{fig:demo}) and vary $n_{\ce{TiO2}}/n$ and $T_0$. For the bottom plot, we held $n_{\ce{TiO2}}/n = 10^{-17}$ (see Fig.~\ref{fig:atmo.ntio2}). We computed $J_{*,1}$ using perturbed values of temperature $T_1$ and pressure $p_1$. We obtained those values using the adiabatic equations of state, assuming a density perturbation resulting from the passage of an internal gravity wave in the most favourable scenario $\rho_1/\rho_0 = \pm 0.48$ as shown in Fig.~\ref{fig:amp-dust}. We note that as a result of the wide exploration of parameter space, not all points in Fig.~\ref{fig:param-space} correspond to a self-consistently calculated $pT$ atmospheric equilibrium state.

Figure~\ref{fig:param-space} shows that the increase in nucleation rate $J_{*,1}$ makes up a larger portion of the total nucleation rate at higher temperatures, lower background pressures (top plot), and lower $n_{\ce{TiO2}}/n$ (bottom plot). These conditions are unfavourable to efficient nucleation, resulting in very small values of $J_{*,0}$. However, the passage of an interval gravity wave can produce temporary, localised conditions allowing nucleation to occur at an enhanced rate. While that increase in nucleation may not be large in absolute terms, it is much larger than the background values, and leads to a large relative increase. This is consistent with results obtained by \citet{Helling_2001} for simulated sound waves.

To elucidate this point further, we consider a slice of constant $n_{\ce{TiO2}}/n$ in the top plot of Fig.~\ref{fig:param-space}. As the temperature increases  the equilibrium nucleation rate decreases which inhibits the growth of \ce{TiO2} clusters since nucleation favours cooler temperatures. As a result, the nucleation rate enhancement $J_{*,1}$ due to the passage of an internal gravity wave relative to the equilibrium $J_{*,0}$ is diminished leading to an increase in $J_{*,1}/J_{*}$. The opposite occurs if the temperature decreases, leading to an increase in $J_{*,1}/J_{*}$. A similar response is evident when considering a slice of constant $p_0$ in the bottom plot of Fig.~\ref{fig:param-space}.

Further to this, we consider a slice of constant background temperature in the top plot of Fig.~\ref{fig:param-space}. As $n_{\ce{TiO2}}/n$ increases, the equilibrium nucleation rate increases due to the increased supersaturation ratio $S$, since the gas-phase \ce{TiO2} molecules are more likely to cluster and nucleate. As a result, $J_{*,1}$ is diminished relative to the equilibrium $J_{*, 0}$, leading to a decrease in $J_{*,1}/J_{*}$. Similarly, consider a constant slice of temperature $T$ in the bottom plot of Fig.~\ref{fig:param-space}: increasing the background pressure leads to a greater absolute \ce{TiO2} density $n_{\ce{TiO2}}$, resulting in an increased background nucleation rate and lower relative increase due to the passage of an internal gravity wave $J_{*,1}/J_{*}$. If $n_{\ce{TiO2}}/n$ decreases the opposite case occurs, leading to an increase in $J_{*,1}/J_{*}$.

To place the relative nucleation rate $J_{*,1}/J_{*}$ in a wider context, in the bottom plot of Fig.~\ref{fig:param-space} we overplot the $pT$ profiles for the brown dwarf model used for the numerical simulations and of typical and low-gravity L and T dwarfs (see Fig.~\ref{fig:atmo.profile} and Table~\ref{table:atmo.models}). Furthermore, in the top plot of Fig.~\ref{fig:param-space} we plot the line of constant temperature corresponding to an atmospheric pressure of $p_{0}\approx \SI{e-5}{\bar}$ for each of the model atmospheres considered. We note that each model profile has self-consistently calculated nucleation rate as a function of atmospheric pressure $p_{0}$, temperature $T_{0}$, $n_{\ce{TiO2}}$ and atmospheric chemistry that may not correspond to a singular point on the parameter space contours. Therefore, these profiles give a helpful indication of the impact of internal gravity waves on the nucleation rate for the variety of sub-stellar objects considered. For example, we consider the solid line ($T_\textnormal{eff}=\SI{1500}{\K}, \log g = 5.0$) in the top plot of Fig.~\ref{fig:param-space}. For $n_{\ce{TiO2}}/n = 10^{-17}$, we can deduce $J_{*,1}/J_{*} \approx 0.95$, equivalent to $J_{*,1} \approx 20 J_{*,0}$, which is consistent with the data shown in Fig.~\ref{fig:atmo.ntio2}. The same result can be obtained using the solid line in the bottom plot, taking $p_0 \approx 4 \times 10^{-5}$. The intersection of the constant pressure line and the model line yields $J_{*,1}/J_{*} \approx 0.95$ ($J_{*,1} \approx 20 J_{*,0}$).

In general, Fig.~\ref{fig:param-space} demonstrates that the impact of internal gravity waves on the nucleation rate is significant across the sub-stellar objects considered. The profiles in Fig.~\ref{fig:param-space} show that the strongest relative increases in nucleation are obtained when the conditions at equilibrium are less suited for efficient nucleation, which leads to any increase caused by the passage of a gravity wave to be comparatively large. This is visible on the top plot, where the warmer L dwarf models (LD: $T_\textnormal{eff} = \SI{2000}{\K}, \log{g}=4.5$; LGLD: $T_\textnormal{eff} = \SI{2000}{\K}, \log{g}=3.0$) are linked to strong increase in nucleation for a wider range of $n_{\ce{TiO2}}/n$ than the cooler T dwarf models (TD: $T_\textnormal{eff} = \SI{1200}{\K}, \log{g}=4.5$; LGTD: $T_\textnormal{eff} = \SI{1200}{\K}, \log{g}=3.0$). Similarly, the $pT$ profiles on the bottom plots show that for a constant $n_{\ce{TiO2}}/n$, the warmer L dwarwillf models exhibit a stronger relative increase in nucleation than T dwarf models.

\section{Discussion}\label{sec:discussion}

This paper has investigated and characterised the effect of linear internal gravity waves on the evolution of dust clouds in sub-stellar (brown dwarf and gas giant exoplanetary) atmospheres for the first time. We have shown that in numerical fluid simulations, the passage of an internal gravity wave leads to an increase of dust nucleation by up to a factor $20$, and an increase of dust mantle growth rate by up to a factor $1.6$. Through an exploration of the wider sub-stellar parameter space, we have shown that, in absolute terms, the increase in dust nucleation due to internal gravity waves is stronger in cooler (T dwarfs) and \ce{TiO2}-rich sub-stellar atmospheres. The relative increase, however, is greater in warm (L dwarf) and \ce{TiO2}-poor atmospheres due to conditions that are less suited for efficient nucleation at equilibrium. This latter point is important since the stronger the contrast between the perturbed and equilibrium values, the better the chance of detecting an observable signal. Recent observations \citep{Marocco_2014} and models \citep{Hiranaka_2016} suggest that the extreme reddening of some L dwarfs could be due to a dust haze layer high up in their atmosphere, which could potentially be impacted by internal gravity waves.

The presence of the signature of an internal gravity wave in the spectra of a brown dwarf could indicate the presence of convection deep in the atmosphere or, in the case of a terrestrial exoplanetary atmosphere, that the body has a rocky, solid surface with relief, since such features are known to trigger the buoyancy oscillation required to generate the waves \citep{Roeten_2019}. In such a scenario, the wavelength of the resulting wave could give an indication of the scale of the perturbing feature. Further investigation into the non-linear evolution of internal gravity waves could potentially yield greater variations in atmospheric density and nucleation rate. Moreover introducing additional effects, such as the Coriolis effect and dynamical equilibria, and investigating their impact on the evolution of internal gravity waves and the resulting cloud cover, might yield further insight into inhomogeneous cloud coverage in sub-stellar atmospheres.


Additionally, observations of of the photometric variability resulting from the propagation of an internal gravity wave could provide a novel way of diagnosing the atmospheric gas density. We consider two identical, adjacent vertical atmospheric profiles. We assume that one profile contains a gas over-density of amplitude $\rho_{1}$, and a corresponding dust over-density $\rho_{d1}$, that occurs over a spatial length scale $L_{0}$, as the result of a propagating internal gravity wave. The ratio of the spectral flux density $S$ from both can be expressed in terms of their respective optical depths,
\begin{equation}
\frac{S_{1}}{S_{0}}\approx\exp{(\tau_{1}-\tau_{0})}
\end{equation}
where
\begin{equation}
\tau_{i}=\int_{0}^{D}\sum_{s}\kappa_{s}\rho_{s}~\textnormal{d}l,
\end{equation}
and we have assumed that the solid angle subtended by the features is the same. In a simple approach, we only consider absorption contributions from the gas and the dust in the atmosphere. To simplify matters further, we assume a total mean opacity to represent the contributions from the gas $\kappa_{g}$ and from the dust $\kappa_{d}$ respectively. The non-zero contributions of the optical depth integral over the extent of the atmosphere $D$, can be approximated to give,
\begin{equation}
\frac{S_{0}}{S_{1}}\approx\exp{[L_{0}(\kappa_{g}\rho_{1}+\kappa_{d}\rho_{d1})]}.
\end{equation}
In the linear regime, Figure~\ref{fig:amp-dust} maps the normalised amplitude of the driving density perturbation, $\rho_{A}/\rho_{0}$, to the normalised amplitude of the resulting density, $\rho_{1}/\rho_{0}$, and nucleation rate, $J_{*1}/J_{*0}$, wave response,
\begin{align}
\frac{\rho_{1}}{\rho_{0}}&=\chi\frac{\rho_{A}}{\rho_{0}}, \\
\frac{J_{*1}}{J_{*0}}&=\delta\frac{\rho_{A}}{\rho_{0}},
\end{align}
$\chi$ and $\delta$ are functions of $\rho_{A}/\rho_{0}$. Therefore, the ratio of flux densities can be expressed as,
\begin{equation}
\frac{S_{0}}{S_{1}}\approx\exp{[L_{0}(\chi\kappa_{g}\rho_{A}+\kappa_{d}\rho_{d1})]} \label{step1}
\end{equation}
where
\begin{align}
\rho_{d1}&=n_{d1}m_{d}, \nonumber \\
&\approx J_{*1}\Delta t m_{d} ,\nonumber \\
&\approx J_{*0}\delta\frac{\rho_{A}}{\rho_{0}}\Delta t m_{d},
\end{align}
where $m_{d}=4\pi a_{d}^{3}\rho_{m}/3$ is the mass of a dust particle, and $a_{d}$ is the radius of a dust grain. To relate the amplitude of the initial density perturbation to the equilibrium atmospheric density, we solve the differential equation for the buoyancy frequency (Eq.~(\ref{eqn:bv-freq})) over the length scale of the perturbation $L_{0}$, giving
\begin{equation}
\rho_{A}\approx\rho_{0}\exp{(- qN^{2}L_{0}/g)},
\end{equation}
where we have assumed that the $N$ is approximately constant across $L_{0}$, and $qN$ is the wave frequency with $q\in(0,1]$ (we assume $q=1$ for simplicity). Therefore, rearranging Eq.~(\ref{step1}), we obtain an expression for the equilibrium atmospheric density:
\begin{equation}
\rho_{0}\approx\frac{\exp{(N^{2}L_{0}/g)}}{L_{0}\kappa_{g}\chi}[\ln{(S_{0}/S_{1})}-L_{0}\kappa_{d}\rho_{d1}],
\end{equation}
where
\begin{equation}
\rho_{d1}\approx J_{*0}\delta\Delta t m_{d}\exp{(-N^{2}L_{0}/g)}. \label{obs}
\end{equation}
This expression allows us to estimate the density of a sub-stellar atmosphere based on potential observations of $S_{0}/S_{1}$ on a timescale consistent with the buoyancy frequency. As an example, for the order of magnitude values list in Table~\ref{table:an.params}, Eq.~(\ref{obs}) gives an estimation of $\rho_{0}\approx \SI{e-5}{\kg\per\m\cubed}$, which is consistent with contemporary atmospheric numerical models. This demonstrates that from observations of $S_{0}/S_{1}$, $L_{0}$, and the timescale of variation, an estimation of the atmospheric density of a sub-stellar atmosphere can be made. A more in-depth analysis could involve calculating synthetic spectra showing the impact of the gravity wave that could be expected from observations, and will be considered in a further paper.

\begin{table*}
\caption{Example order of magnitude values used in Eq.~(\ref{obs})}
\label{table:an.params}
\centering
\begin{tabular}{l l l}
\hline\hline
Parameter & Value &  Notes \\
\hline
$L_{0}$ & $10^{3}$~m  & See \citet{Freytag_2010,Marley_2015}\\
$\kappa_{g}$ & \SI{e-2}{\m\squared\kg} & See Fig. 13, \citet{Lee_2016} for $\lambda\approx \SI{1}{\micro\m}$ \\
$\kappa_{d}$ & \SI{e-4}{\m\squared\kg} & See Fig. 13, \citet{Lee_2016} for $\lambda\approx \SI{1}{\micro\m}$ \\
$\Delta t$ & $=T_{N}=\SI{10}{\s}$  & See Fig.~\ref{fig:atmo.period}\\
$\chi$ & $2$ & Fig.~\ref{fig:amp-dust} for $\rho_{A}/\rho_{0}=0.1$, $\omega/N=0.25$\\
$\delta$ & $30$ & Fig.~\ref{fig:amp-dust} for $\rho_{A}/\rho_{0}=0.1$, $\omega/N=0.25$\\
$J_{*0}$ & \SI{e4}{\per\m\cubed\per\s} & Eq.~(\ref{eqn:nucleation}) at equilibrium, see also Fig.~1 in \citet{Helling_2008b}\\
$S_{0}/S_{1}$ & $1.001$  & see Figs~7$-$10 in \citet{Buenzli_2014}\\
$m_{d}$ & \SI{e-15}{\kg} & for $a_{d}=\SI{e-6}{\m}$, $\rho=\SI{e3}{\kg\per\m\cubed}$ \\
$g$ & \SI{e3}{\m\per\s\squared} & $\log{g}=5$ \\
$N^{2}$ & $=4\pi^{2}/\Delta t$ &  \\
\hline
\end{tabular}
\end{table*}

\begin{acknowledgements}
The authors are grateful to the anonymous referee for constructive comments and suggestions that have improved this paper.

A.P. is grateful for funding and support received from Abertay University as part of the RLINCS studentship programme.

C.R.S. is grateful for funding from the Royal Society via grant number RG160840 and from the Carnegie Trust for the Universities of Scotland via research incentive grant number RIG007788.

E.K.H. Lee acknowledges support from the University of Oxford and CSH Bern through the Bernoulli fellowship and support from the European community through the ERC advanced grant project EXOCONDENSE (PI: R.T. Pierrehumbert).
\end{acknowledgements}

\bibliographystyle{aa}
\bibliography{parent_2019}
\end{document}